\documentclass[aps,prb,epsf,twocolumn,showpacs]{revtex4-2}

\usepackage{CJK,amsmath,amssymb,graphics,epsfig,epstopdf,color,verbatim,ulem,braket,tabularx,float,makecell,multirow}
\usepackage[colorlinks,linkcolor=blue,citecolor=blue,urlcolor=blue]{hyperref}

\begin{document}
\title{Ground-state phase diagram and Haldane-like SPT phase in the $S=1/2$ Heisenberg model on the square-hexagon-octagon lattice}

\author{Yumeng Luo}
\author{Yuehong Li}
\author{Mengfan Jiang}
\author{Muwei Wu}
\author{Jian-Jian Yang}
\author{Dao-Xin Yao}
\email{yaodaox@mail.sysu.edu.cn}
\author{Han-Qing Wu}
\email{wuhanq3@mail.sysu.edu.cn}
\affiliation{Center for Neutron Science and Technology, \\Guangdong Provincial Key Laboratory of Magnetoelectric Physics and Devices, \\State Key Laboratory of Optoelectronic Materials and Technologies,\\
\mbox{School of Physics, Sun Yat-sen University, Guangzhou 510275, China.}}

\date{\today}

\begin{abstract}
Using stochastic series expansion quantum Monte Carlo and density matrix renormalization group methods, we investigate the ground-state phase diagram of the $S=1/2$ Heisenberg model on the two-dimensional square-hexagon-octagon (SHO) lattice. The model incorporates nearest-neighbor interactions $J_1$ (intrahexagon interaction) and $J_2$ (interhexagon), as well as a selected third-neighbor interaction $J_3$ along the $x$ direction. We identify five distinct phases in the parameter regime $0<\lambda_1=J_2/J_1<4, 0<\lambda_2=J_3/J_1<4$: a N\'eel antiferromagentic phase, two dimer phases (orthogonal and ladder staggered dimers), a hexagon singlet phase, and notably a Haldane-like symmetry-protected topological (SPT) phase. The topological nature of the Haldane-like phase is confirmed by the degeneracy of the ground-state energy under open boundary conditions and the twofold degeneracy of the entanglement spectrum. Phase boundaries are accurately determined using finite-size scaling of the spin stiffness and Binder cumulant. Data collapse analysis reveals that all transitions from nonmagnetic phases to the antiferromagnetic phase belong to the three-dimensional $O(3)$ Heisenberg universality class. In addition, we investigate the robustness of the SPT phase to other interactions, such as those that could arise in experimental materials. Our work establishes a comprehensive theoretical framework for understanding magnetic and topological phases on the SHO lattice.
\end{abstract}
\maketitle

\section{Introduction}
\begin{figure}
\centering
\includegraphics[width=1\linewidth]{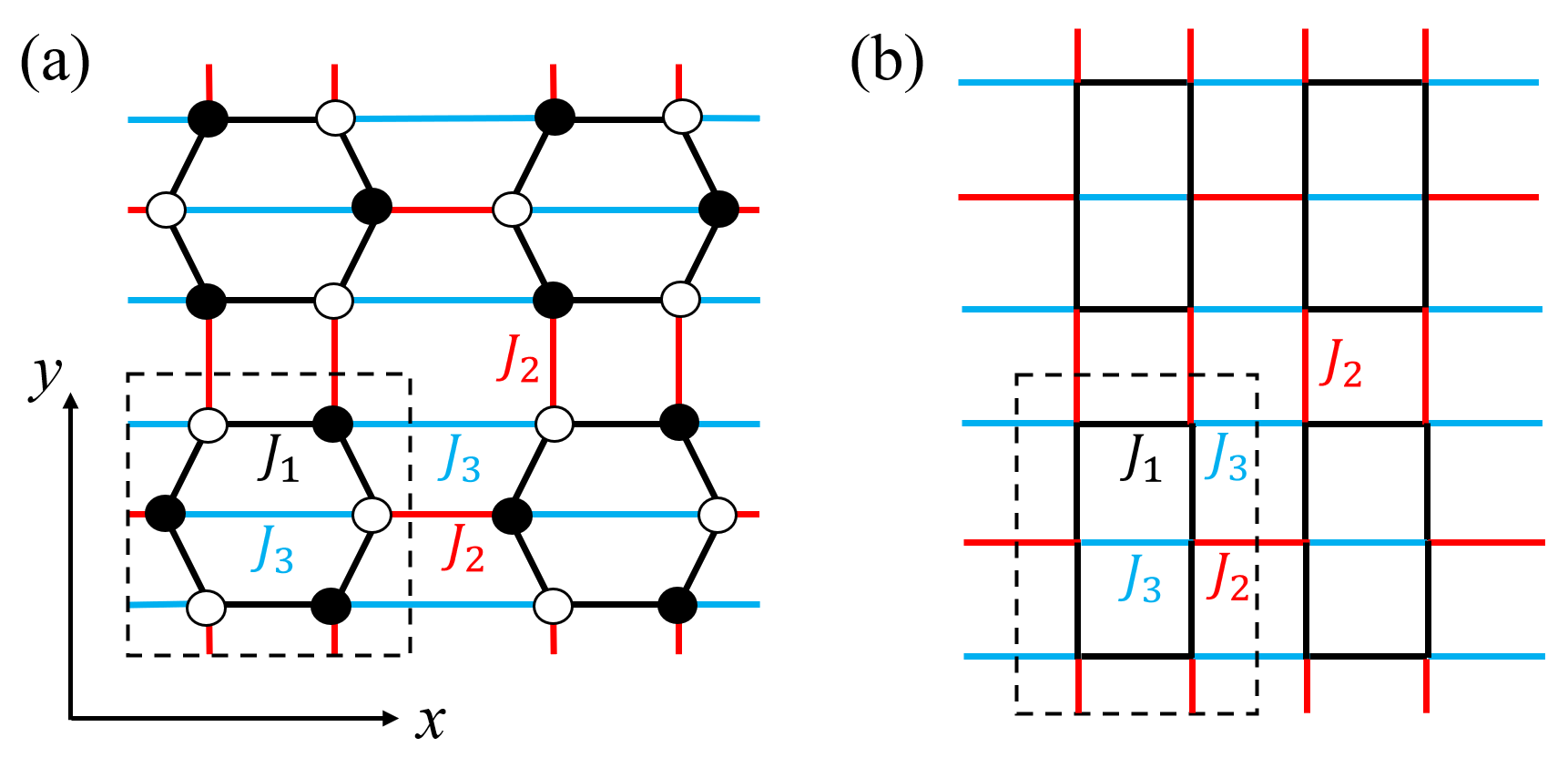}
\caption{\label{fig:FIG.1.}(a) The sketch of the SHO structure, constituted with hexagon unit cells. There are three types of bonds with different interactions from our definition: $J_1$ is indicated with black color, $J_2$ is red, and $J_3$ is blue. We also define the ratios between them: $\lambda_1=J_2/J_1$ and $\lambda_2=J_3/J_1$. A unit cell of the SHO lattice consists of six sites, which is shown in the black dashed square. (b) The lattice in square form is topologically equivalent to the SHO lattice.}
\end{figure}

Since the mechanical exfoliation of a single layer of graphene in 2004~\cite{doi:10.1126/science.1102896,RevModPhys.83.837}, two-dimensional (2D) material systems have garnered significant attention~\cite{RevModPhys.81.109,novoselovRoadmapGraphene2012,030a37cb2125447faf8c34ab0a4904b9,huangLayerdependentFerromagnetismVan2017}. These breakthroughs has spurred extensive research into the properties and applications of 2D materials. Graphene is a periodic hexagonal network built from $sp^2$-hybridized carbon atoms, with excellent electrical and mechanical properties. Based on the structure of graphene, researchers try to obtain two-dimensional materials with some new properties by adding nonhexagons to replace six-membered rings in graphene. For example, a new kind of graphenelike nanoribbons periodically embedded with four- and eight-membered rings has been generated~\cite{liuGraphenelikeNanoribbonsPeriodically2017}. In this material, the nonhexagonal rings alter the electronic properties of the nanoribbons due to the topological change. Similar to graphene nanoribbons, 2D nonbenzenoid carbon allotropes have also attracted great interest from researchers. In 2021, an article reported a 2D biphenylene network composed of $sp^2$-hybridized carbon atoms with periodically arranged four-, six-, and eight-membered rings~\cite{doi:10.1126/science.abg4509}, achieving a two-dimensional square-hexagon-octagon (SHO) lattice, as illustrated in Fig.~\ref{fig:FIG.1.}. The SHO lattice as a new kind of unfrustrated 2D lattice offers a new platform for studying electronic correlations, magnetism, and superconducting properties.

The electronic band structures were systematically investigated using density functional theory~\cite{LIU2022153993,Ye_2023,hudspethElectronicPropertiesBiphenylene2010}, and the possible superconductivity properties of biphenylene were studied using the random phase approximation~\cite{Ye_2023}. Furthermore, the ground-state phase diagram and critical behavior of the Hubbard model on the SHO lattice were examined using the determinant quantum Monte Carlo method~\cite{PhysRevB.109.155122}. Additionally, the topological and electronic properties on an anisotropic SHO lattice were also studied~\cite{PhysRevB.108.144407}. However, the magnetic properties of the Heisenberg model on the SHO lattice remain an open question. It is of great interest to investigate the ground-state properties of the $S=1/2$ Heisenberg model on the SHO lattice. Although no SHO lattice magnetic materials have been identified in transition metals thus far, there is potential for their discovery in the future. Our research offers crucial theoretical analysis in anticipation of such findings. Moreover, magnetic systems of this type could potentially be realized in experimental platforms such as cold atoms or metal-organic frameworks. As a new theoretical model, it is of significant value for the systematic investigation of its rich phase diagram and critical behavior.

In this work, we employ the stochastic series expansion (SSE) quantum Monte Carlo (QMC)~\cite{10.1063/1.3518900,Sandvik1991,Syljuasen2002,PhysRevB.56.11678} and density matrix renormalization group (DMRG)~\cite{PhysRevLett.69.2863,PhysRevB.48.10345,RevModPhys.77.259,Schollwoeck:2010uqf,Ostlund1995,XWang1997} methods to study the ground-state phase diagram of the $S=1/2$ Heisenberg model on the SHO lattice. To enrich the phase diagram and enhance its topological equivalence to the square lattice, we introduce the selected third-nearest-neighbor interaction $J_3$ along the $x$ direction. Notably, among the discovered phases—which include N\'eel, dimerized, and hexagon phases—we identify a Haldane-like phase~\cite{PhysRevLett.59.799,PhysRevLett.50.1153}. This phase has a unique ground state which does not break any symmetries and exhibits a finite triplet excitation gap. Its topological properties can be evidenced by the degeneracy of the ground-state energy levels under open boundary conditions (OBCs) and the twofold degeneracy of the entanglement spectrum. This phase is distinct and cannot be continuously
connected to trivial phase in the presence of
certain symmetries, making it a symmetry-protected topological (SPT) phase~\cite{PhysRevB.80.155131,PhysRevB.81.064439,PhysRevB.83.035107,PhysRevB.87.155114}. Similar phases have been observed in $S=1/2$ two-leg ladder systems~\cite{PhysRevB.73.214427,PhysRevB.86.195122,PhysRevLett.93.177004}. Furthermore, to establish connections with potential experimental material synthesis, we also investigate the phase diagram of this SPT phase with respect to other possible interactions. To obtain accurate phase boundaries and critical behaviors between different phases, we employ comprehensive finite-size scaling techniques~\cite{PhysRevE.88.061301,PhysRevB.85.014414} to verify the universality class.

The paper is organized as follows. In Sec.~\ref{MODEL AND METHODS}, we introduce the model Hamiltonian and outline the physical quantities used to characterize the ground-state phase diagram of the SHO lattice. Then, in Sec.~\ref{NUMERICAL RESULTS}, we employ SSE-QMC and DMRG methods combined with finite-size scaling to numerically determine the accurate phase diagram. Specifically, Sec.~\ref{A} examines the case with $\lambda_2=J_3/J_1=0$, Sec.~\ref{B} considers the scenario where $\lambda_1=J_2/J_1=0$, and Sec.~\ref{C} explores the full phase diagram. Following that, Sec.~\ref{D} focuses on the critical behavior of phase transitions on the SHO lattice. Moreover, Sec.~\ref{E} investigates the stability of the SPT phase against other interactions. Finally, in Sec.~\ref{CONCLUSIONS}, we summarize all identified ground-state phases and provide a concise discussion.

\section{MODEL AND METHODS\label{MODEL AND METHODS}}
The $S = 1/2$ Heisenberg model on the SHO lattice is described by the Hamiltonian:
\begin{align}
H=&J_1\sum_{\langle i,j\rangle}\mathbf{S}_i\cdot\mathbf{S}_j+J_2\sum_{\langle i,j\rangle^\prime}\mathbf{S}_i\cdot\mathbf{S}_j\notag \\
&+J_3\sum_{\langle\langle\langle i,j\rangle\rangle\rangle}\mathbf{S}_i\cdot\mathbf{S}_j.
\end{align}
Note that the SHO lattice and its topologically equivalent squarelike lattice are depicted in Figs.~\ref{fig:FIG.1.}(a) and \ref{fig:FIG.1.}(b), respectively. The Hamiltonian includes several types of spin-spin interactions: $\langle i,j\rangle$ and $\langle i,j\rangle'$ represent intra- and interhexagon nearest-neighbor couplings, respectively, while $\langle\langle\langle i,j\rangle\rangle\rangle$ corresponds to the selected third-nearest-neighbor interaction along the $x$ direction. This $J_3$ term also acts as a nearest-neighbor interaction in the topologically equivalent squarelike lattice. Here, $S_i$ denotes the spin operator at site $i$. For simplicity, we assume isotropic interhexagon interactions, meaning that the horizontal and vertical $J_2$ bonds are set equal. The $J_3$ coupling was introduced primarily to establish a structural connection with the conventional square-lattice Heisenberg model. To facilitate comparison of interaction strengths, we define the dimensionless ratios $\lambda_1 = J_2/J_1$ and $\lambda_2 = J_3/J_1$.

In the following text, unless specified, we generally employ periodic boundary conditions (PBCs) in QMC calculations. The linear system size is denoted as $L$, and the total number of grid points is given by $N=6L^2$. In the SSE-QMC calculation, we use the staggered magnetization along the $S^z$ direction and the squared staggered magnetization $M_s^2$ to illustrate the N\'eel order:
\begin{align}
M_s^z=&\frac{1}{N}\sum_{i=1}^{N}\phi_i\langle S_i^z\rangle,\\
M_s^2=&\frac{1}{N^2}\sum_{ij}\phi_i\phi_j\langle\mathbf{S}_i\cdot\mathbf{S}_j\rangle.
%=3\left(M_s^z\right)^2.
\end{align}
The sign factor $\phi_i$ denotes the $+1$ or $-1$ sign at site $i$ for different sublattices, which is illustrated as staggered black and white circles in Fig.~\ref{fig:FIG.1.}(a). We also use the following physical quantities that help to obtain the accurate phase boundaries: spin stiffness and Binder cumulant,
\begin{align}
\rho_s^a&=\left.\frac{1}{N}\frac{\partial^2E_0(\theta)}{\partial\theta^2}\right|_{\theta=0}\\
&=\frac3{2\beta N}\left\langle\left(N_a^+-N_a^-\right)^2\right\rangle,\quad a=x,y,\\
U_2&=\frac{3}{2}\left(1-\frac{1}{3}\frac{\langle (M_s^z)^4\rangle}{\langle (M_s^z)^2\rangle^2}\right).
\end{align}
Spin stiffness denotes the ground-state energy cost of applying a twisting angle $\theta$ to the rotors of a spin system, which can be calculated by the number of $S_i^+S_j^-$ and $S_i^-S_j^+$ operators in the SSE program. Here $E_0$ is the ground-state energy of the twisted 
Hamiltonian, $N$ is the number of spins in the system, $N_a^+$ is the number of $S_i^+S_j^-$ operators, and $N_a^-$ is the number of $S_i^-S_j^+$ operators. The value of $\beta=1/T$ is taken to be proportional to the system size $L$ when doing the scaling. The factor $3/2$ in the Binder cumulant is for normalization, i.e., $U_2\rightarrow1$ when the system is a magnetic ordered state and $U_2\rightarrow0$ when it is a disordered state, for $N\rightarrow\infty$.

When $J_2=0$, the system becomes a ladder system. To analyze the ground-state phases in this scenario, we use the DMRG method to compute the energy gaps and entanglement spectra. In our DMRG calculations, we use 4000 $SU(2)$ symmetric states to ensure that the density matrix truncation error is below $10^{-8}$. The energy gaps are defined as follows: The singlet gap $\Delta_{S}$ is the energy difference between the first excited state and the ground state in the $S=0$ sector. The triplet gap $\Delta_{T}$ is the energy difference between the lowest $S=1$ state and the $S=0$ ground state. The quintuplet gap $\Delta_{Q}$ is the energy difference between the lowest $S=2$ state and the $S=0$ ground state, i.e.,
\begin{align}
\Delta_{S}=E_1(S=0)-E_0(S=0),
\end{align}
\begin{align}
\Delta_{T}=E_0(S=1)-E_0(S=0),
\end{align}
\begin{align}
\Delta_{Q}=E_0(S=2)-E_0(S=0).
\end{align}
Then we use the finite-size gaps to do the extrapolation to get the energy gaps at the thermodynamic limit.

\section{NUMERICAL RESULTS\label{NUMERICAL RESULTS}}
\begin{figure}
\centering
\includegraphics[width=1\linewidth]{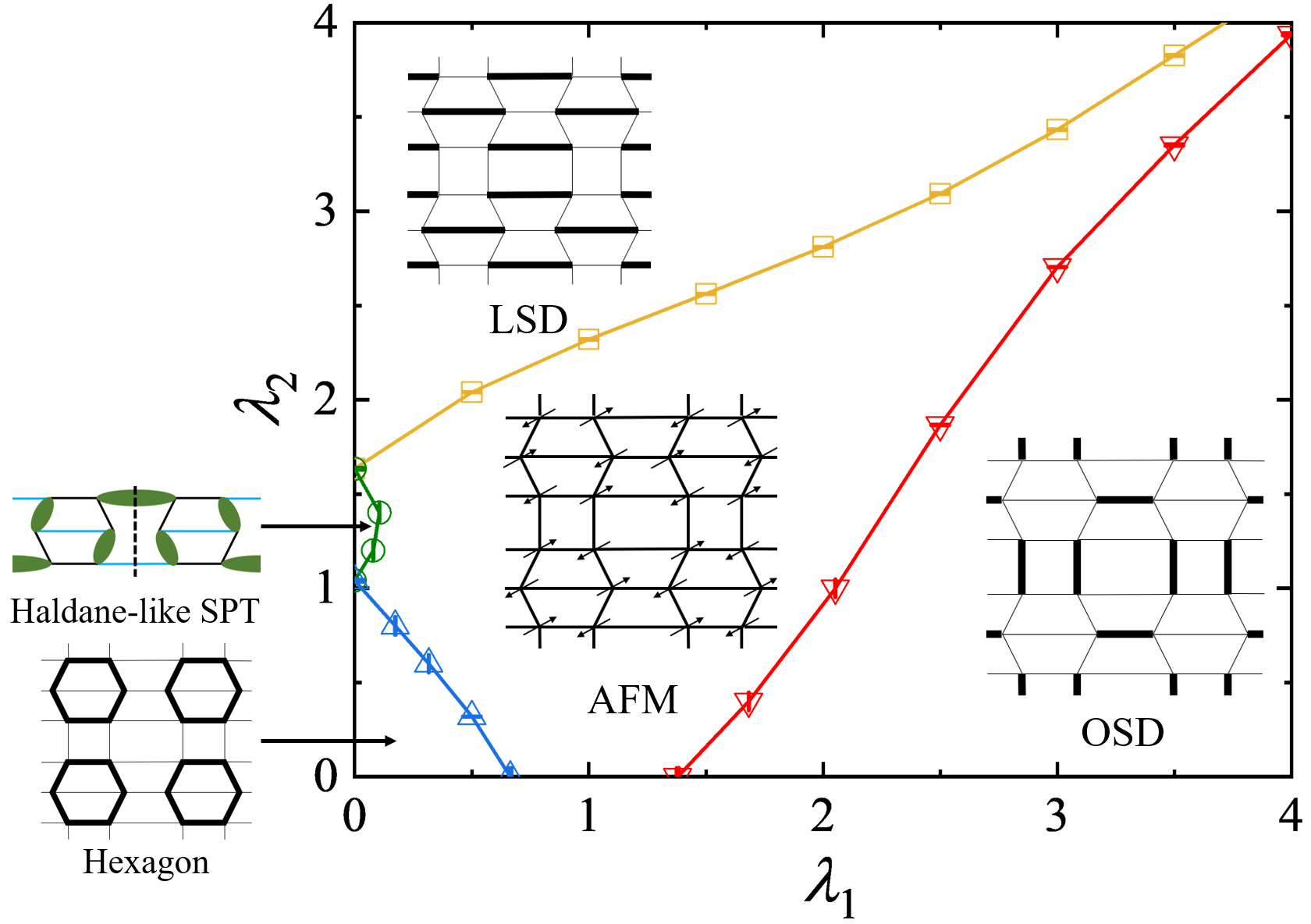}
\caption{\label{fig:FIG.2.}The ground-state phase diagram of the $S = 1/2$ Heisenberg model on the 2D SHO lattice. In total there are five phases in the phase diagram, namely the hexagon phase, the orthogonal staggered dimer (OSD) phase, the ladder staggered dimer (LSD) phase, the Haldane-like SPT phase, and the N\'{e}el antiferromagnetic (AFM) phase.}
\end{figure}

Using SSE-QMC method and finite-size scaling, we obtain the accurate ground-state phase diagram of the $S = 1/2$ Heisenberg model on the 2D SHO lattice, which is shown in Fig.~\ref{fig:FIG.2.}. There are four gapped phases, namely the hexagon phase, the orthogonal staggered dimer (OSD) phase, the ladder staggered dimer (LSD) phase, and the Haldane-like SPT phase, surrounding the gapless N\'{e}el antiferromagnetic (AFM) phase. In the subsequent subsection, we will elaborate on the method used to obtain the phase diagram and provide a detailed overview of the distinct phases.

\subsection{$J_1\text{-}J_2$ ground-state phase diagram\label{A}}
First, we investigate the ground state of the system with $\lambda_2=J_3/J_1=0$ fixed and a changing $\lambda_1=J_2/J_1$. Initially, we examined the ground state of the Heisenberg model on the SHO lattice at two extremes: $\lambda_1$ approaching zero and positive infinity. At $\lambda_1=\lambda_2=0$, the model reduces to isolated, orderly hexagonal lattices. The ground state here is a direct product of hexagonal singlet states with finite local excitation gaps. After adding weak interhexagon interactions, the system is in a hexagon phase with gapped lowest triplet excitation. When $\lambda_1\to\infty$ and $\lambda_2 = 0$, the ground state adiabatically connects to a vertically and horizontally regular dimer product state, termed the orthogonal staggered dimer phase. 

\begin{figure}[t]
\centering
\includegraphics[width=0.5\textwidth]{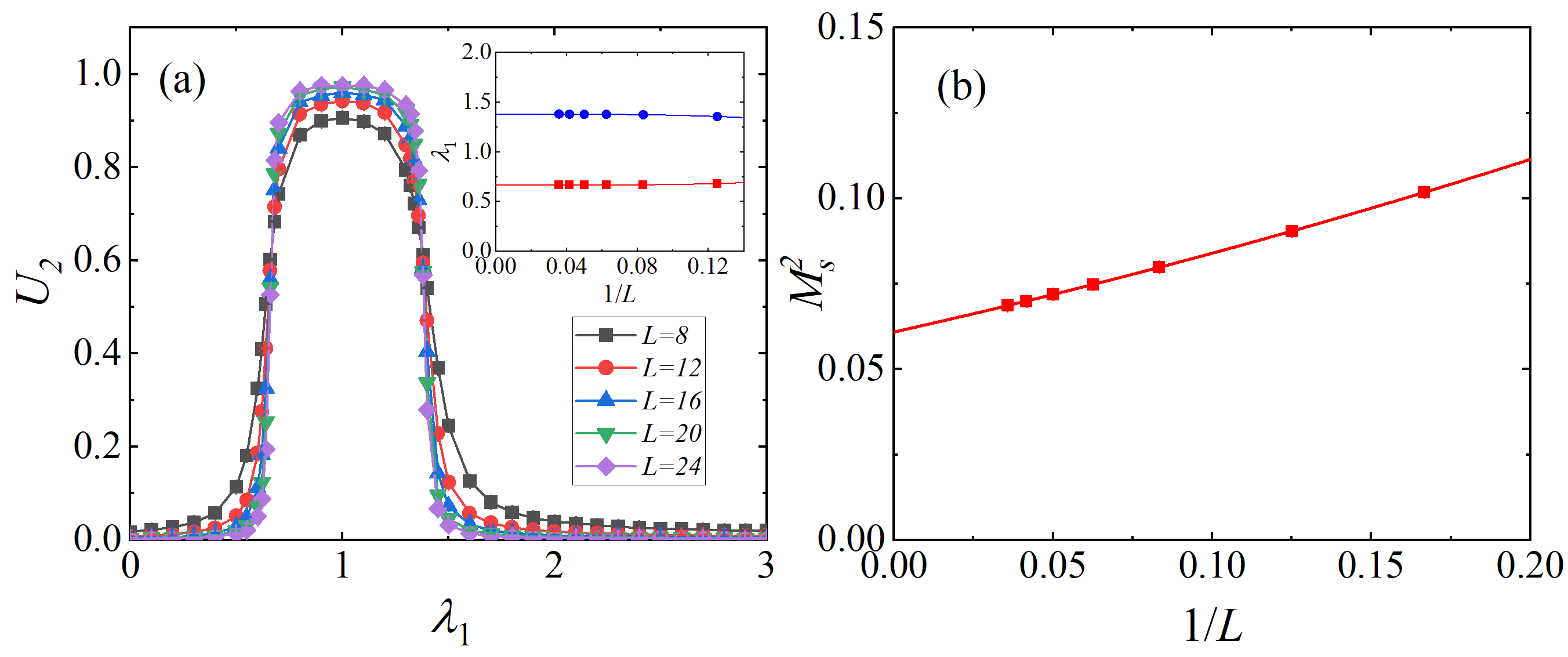}
\caption{\label{fig:FIG.3.} (a) Variation of Binder cumulant $U_2$ with $\lambda_1$ at different sizes when $\lambda_2 = 0$. The inset in (a) is the extrapolation of the two intersection positions of the Binder cumulant as a function of the size, using the extrapolation function in the form of $\lambda_{1,c}(L)=\lambda_{1,c}(\infty)+aL^{-\omega}$~\cite{10.1063/1.3518900}. (b) Second-order polynomial extrapolation of $M_s^2$ as a function of $L$ when $\lambda_1 = 1$. The nonzero extrapolated value indicates that there is a nonzero staggered magnetic order in the thermodynamic limit: $M^2_s(L\to\infty) = 0.0609(2)$, and $SU(2)$ continuous symmetry breaking occurs at the intermediate phase.}
\end{figure}

Regarding the phase between the hexagon phase and the OSD phase, given the even-number sublattice structure of the SHO lattice, it is possible that a long-range magnetic order exists. And it is tentatively speculated to be the N\'{e}el AFM phase. By extrapolating the squared magnetization at $\lambda_1=1$ [as shown in Fig.~\ref{fig:FIG.3.} (b)], we can confirm that the intermediate phase is indeed a Néel AFM phase. To characterize the phase boundaries among the three phases, we show the intersection of Binder cumulant $U_2$ for different sizes when $\lambda_1 \in [0,3]$, which is illustrated in Fig.~\ref{fig:FIG.3.} (a). To pinpoint the crossing location precisely, we computed additional, denser data points in the vicinity of the intersection, enabling a precise determination of the crossing point, as shown in the inset of Fig.~\ref{fig:FIG.3.} (a). Subsequently we perform an extrapolation of this intersection to the thermodynamic limit of $1/L\to0$ using scaling function $\lambda_{1,c}(L)=\lambda_{1,c}(\infty)+aL^{-\omega}$~\cite{10.1063/1.3518900}. Based on the finite-size scaling results, it is evident that the system undergoes phase transitions at $\lambda_{1,c} = 0.664(1)$ and $1.379(1)$ in the thermodynamic limit. 

Hence, when $\lambda_2=0$ and $\lambda_1$ varies (i.e., in the $J_1\text{-}J_2$ model), the ground-state phase diagram consists of three phases: the hexagon phase, the AFM phase, and the OSD phase. The hexagon phase and OSD phase are nonmagnetic, featuring nonzero triplet excitation energy gaps. In contrast, the AFM phase is magnetic, characterized by gapless Goldstone mode excitations.

\subsection{$J_1\text{-}J_3$ ground-state phase diagram\label{B}}
Moving forward to the next step, our aim is to investigate the scenario where $\lambda_1=0$ and $\lambda_2=J_3/J_1$ varies. In this case, the system comprises an array of decoupled ladders. Therefore, to analyze this situation, we only need to focus on the $J_1\text{-}J_3$ Heisenberg model defined on a single ladder. Consequently, we turn to the DMRG method as a supplementary numerical approach for this particular case. Similar to the $J_1\text{-}J_2$ case, there are two decoupled limits corresponding to $\lambda_2=0$ and $\infty$, respectively. $\lambda_2=0$ corresponds to decoupled hexagon limit, and in the small $\lambda_2$ region the system is in the hexagon phase. $\lambda_2=\infty$ corresponds to the LSD phase which is adiabatically connected to the direct product of the isolated dimers in staggering order. 
Between these two phases, there may be one or more phases. 

\begin{figure}[t]
\centering
\includegraphics[width=0.5\textwidth]{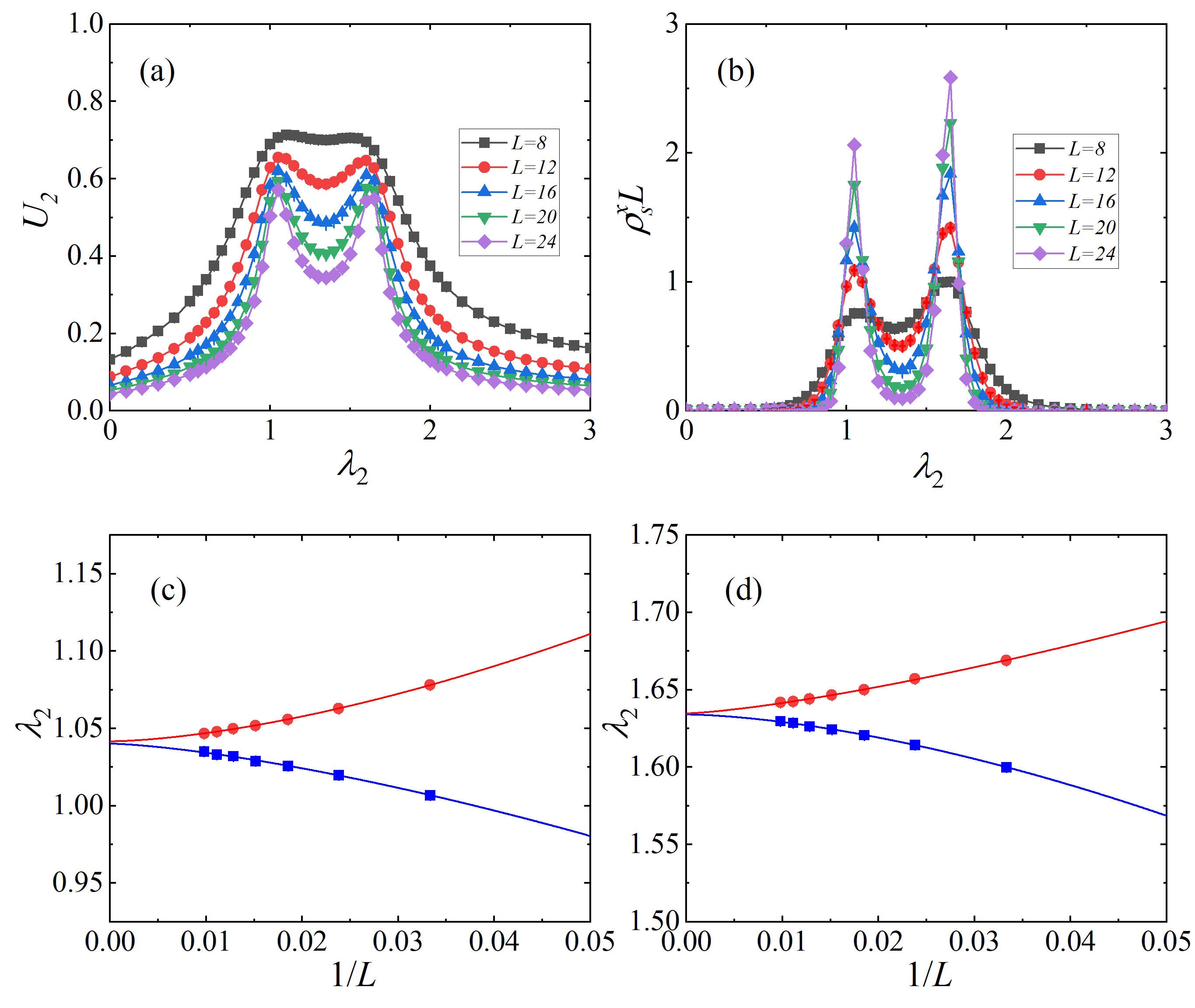}
\caption{\label{fig:FIG.4.}(a) When $\lambda_1 =0$, the Binder cumulant which is defined on a single ladder varies with $\lambda_2$. As the intermediate phase is nonmagnetic, the Binder cumulant fails to reveal phase transitions between nonmagnetic phases, and no intersection points are visible. (b) When $\lambda_1 = 0$, the spin stiffness in the $x$ direction changes with $\lambda_2$. Unlike the Binder cumulant, spin stiffness can detect phase transitions between nonmagnetic phases. However, there are four intersections, so extrapolation is needed to pinpoint the exact phase transition points. (c, d) The extrapolation results for the spin stiffness intersections at the two phase transition points, with the following form of the extrapolation function: $\lambda_{2,c}(L)=\lambda_{2,c}(\infty)+aL^{-\omega}$~\cite{10.1063/1.3518900}. Near $\lambda_{2,c} = 1.040(2)$ and $1.634(2)$, the extrapolation curves approach and intersect each other.} 
\end{figure}

To detect any potential phases between the hexagon phase and the LSD phase, several factors need to be considered. In a quasi-one-dimensional configuration with short-range Heisenberg interaction, quantum fluctuations are relatively strong, preventing the formation of a magnetically ordered phase in the ground state. As a result, the Binder cumulant is not suitable for characterizing phase transitions. Indeed, from the QMC results of the Binder cumulant [Fig.~\ref{fig:FIG.4.}(a)], no intersections are found for different sizes. Therefore, we need to explore other physical quantities to identity the location of the phase transitions. We choose the spin stiffness along the $x$ direction to detect phase transitions, as shown in Fig.~\ref{fig:FIG.4.}(b). The spin stiffness graph shows four intersections between adjacent sizes. To determine whether these four intersections each correspond to a phase transition point, we combine their extrapolation curves for comparison. It turns out that under the thermodynamic limit, two of these four points approach $\lambda_{2,c}=1.040(2)$, and the other two approach $\lambda_{2,c}=1.634(2)$. And these two points are also exactly at the peak positions of spin stiffness. Thus, there are only two phase transition points in the range of $\lambda_1 = 0$, $\lambda_2\in[0,4]$. These two points divide the system into three phases. However, the nature of the middle phase requires further investigation.

\begin{figure}[t]
\centering
\includegraphics[width=0.5\textwidth]{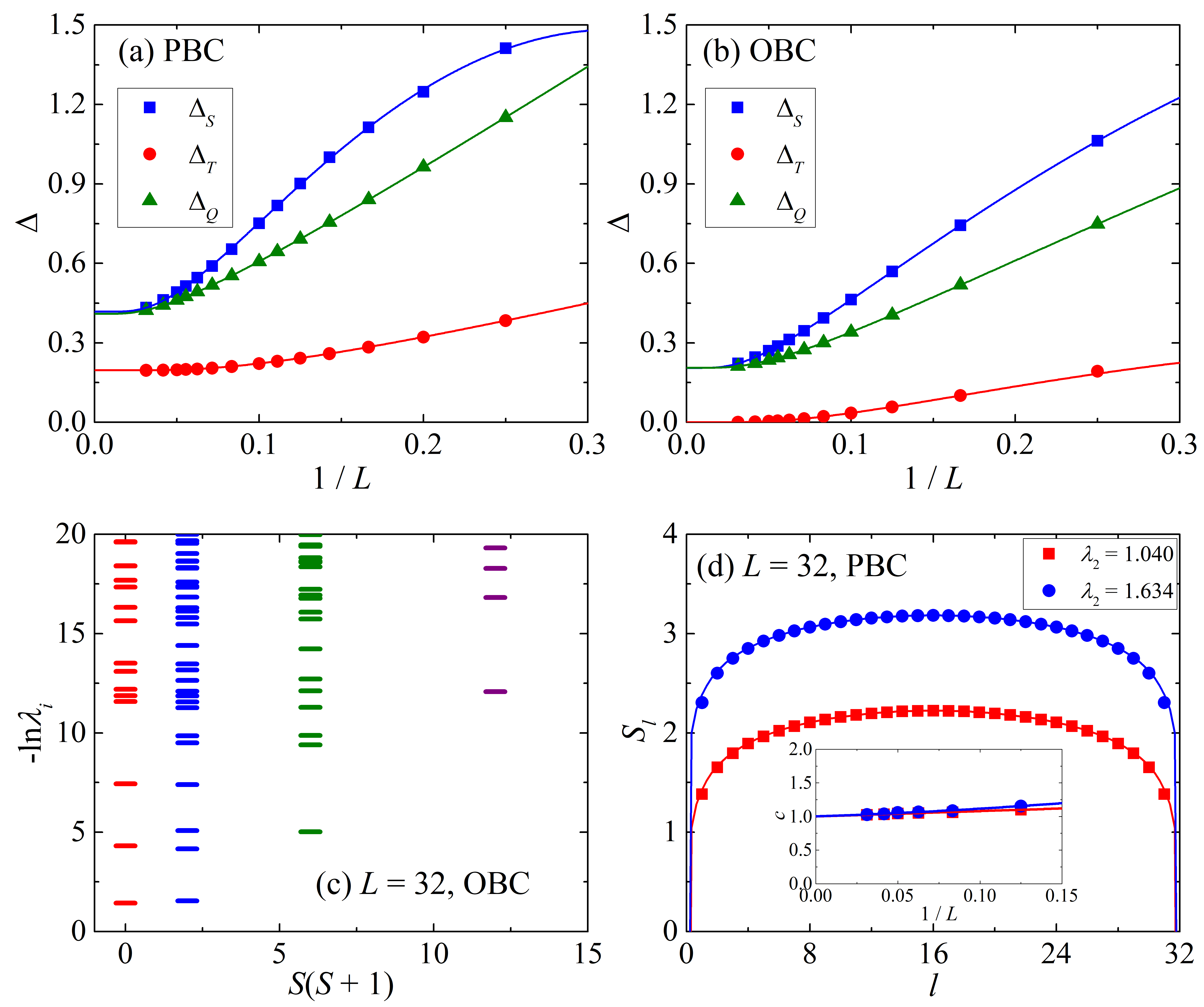}
\caption{\label{fig:FIG.4_3.}(a, b) The extrapolation of the excitation energy gaps for singlet ($S=0$), triplet ($S=1$), and quintuplet ($S=2$) states at $\lambda_2=1.4$, under periodic and open boundary conditions, respectively. The fitting function used is $\Delta(L)=a+e^{-L/\xi}(b/L + c/L^2)$~\cite{Hohenadler2012, JKFang2021}. (c) The entanglement spectrum for a linear system size $L=32$ at $\lambda_2=1.4$ under open boundary conditions. (d) The entanglement entropy near the phase transition points at $\lambda_1=0$, with $L=32$ under periodic boundary conditions. The inset includes linear extrapolations of the central charges $c$ at the these transition points.} 
\label{fig:FIG.5.}
\end{figure}

The intermediate phase cannot be a magnetically ordered phase; instead, it may be a gapless Luttinger liquid or a gapped phase. To determine the excitation energy gaps, we use the density matrix renormalization group method to calculate the energy gaps from the singlet ground state to the excited singlet state, triplet state, and quintuplet state of the finite-size system and then extrapolate to the thermodynamic limit. At $\lambda_2=1.4$, under PBCs, all three gaps extrapolate to finite values [see Fig.~\ref{fig:FIG.5.}(a)]. This indicates that the ground state is unique and fully gapped. In contrast, under OBCs, the first excited triplet state becomes degenerate with the ground state, while the other gaps remain finite, as can be seen in Fig.~\ref{fig:FIG.5.}(b). This suggests the system may be in a bosonic SPT phase with end state under OBCs. To further confirm the topological nature of the intermediate phase, we present the entanglement spectrum in Fig.~\ref{fig:FIG.5.}(c). The entanglement spectrum is computed from the Schmidt decomposition performed at the central bond (see dashed line in the schematic diagram of the SPT phase in Fig.~\ref{fig:FIG.1.}) during the DMRG sweep. From our DMRG calculations, it is evident that all entanglement spectral levels are at least doubly degenerate which is similar to the $S=1$ Haldane chain~\cite{PhysRevB.81.064439}. Hence, the intermediate phase is a Haldane-like SPT phase~\cite{PhysRevB.80.155131,PhysRevB.81.064439}.

%which is protected by any of the following symmetries: $D_2$ symmetry (three $\pi$ spin rotations along three axes), time-reversal symmetry $T$, and mirror symmetries $\sigma_x,\sigma_y$~\cite{PhysRevB.80.155131,PhysRevB.81.064439}. 

Furthermore, in this one-dimensional system, the continuous quantum phase transitions between the Haldane-like SPT phase and the other two phases are effectively captured by the Wess-Zumino-Witten conformal field theory~\cite{moore1989classical}, which provides a framework to understand the universality class of these transitions. The central charge $c$ is a key physical quantity in this context. In our DMRG calculations, we computed the entanglement entropy for different subsystem sizes $l$ while keeping the total system size $L$ fixed [see Fig.~\ref{fig:FIG.5.}(d)]. By fitting the calculated entanglement entropy data under periodic boundary conditions to the Calabrese-Cardy formula $S=\frac{c}{3}ln\left(\frac{L}{\pi}\mathrm{sin}\left(\frac{\pi l}{L}\right)\right)+\mathrm{const}$ [see Fig.~\ref{fig:FIG.4_3.} (d)]~\cite{Pasquale2004, KRen2023}, and then extrapolating to $L=\infty$, we are able to determine the central charges $c=1$ in the thermodynamic limit. 

\begin{figure}[t]
\centering
\includegraphics[width=0.5\textwidth]{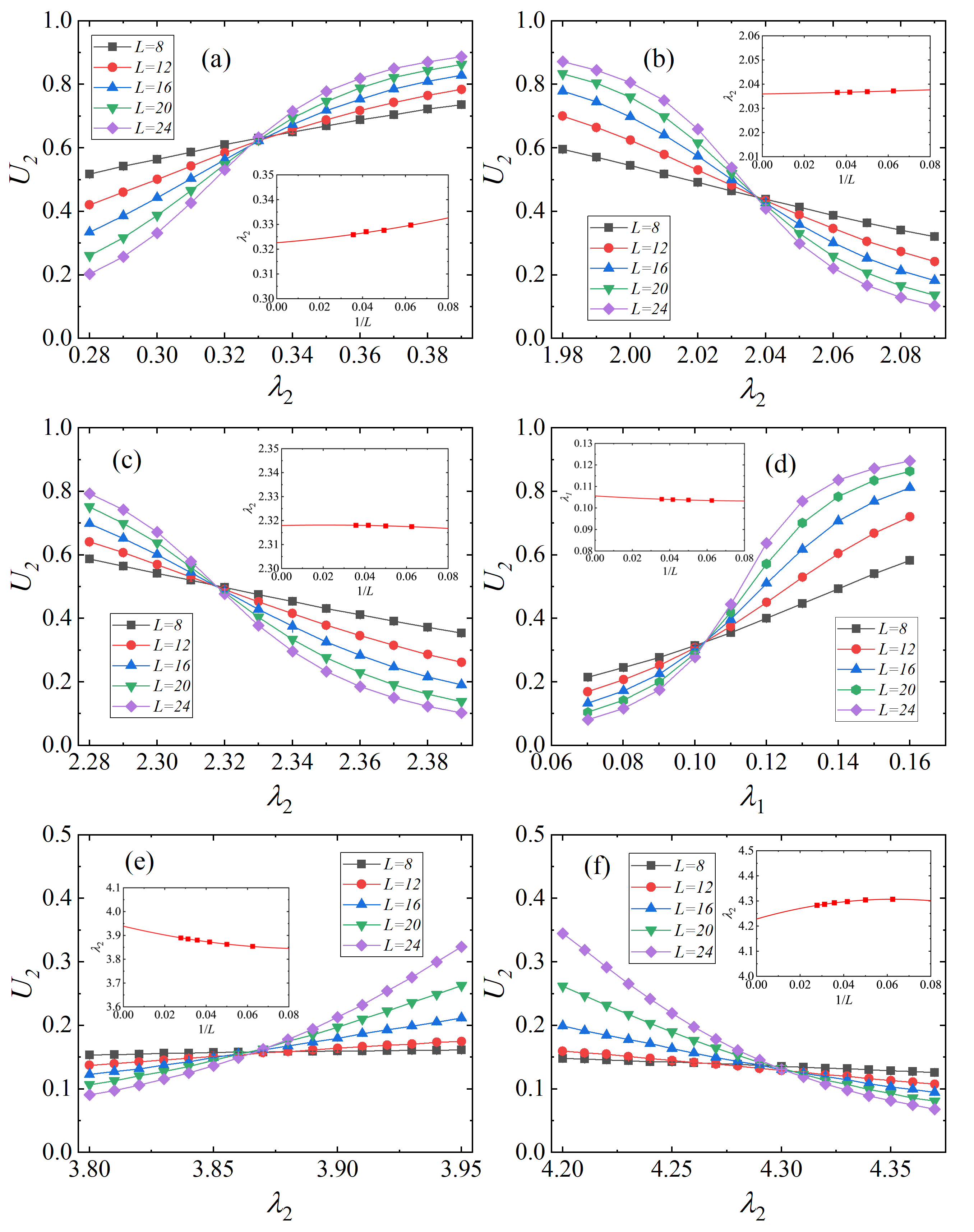}
\caption{\label{fig:FIG.6.}(a, b) Binder cumulant $U_2$ as a function of $\lambda_2$ near the two critical points for $\lambda_1=0.5$. (c) $U_2$ vs $\lambda_2$ near the critical point for $\lambda_1=1$. (d) $U_2$ as a function of $\lambda_1$ near the critical point for $\lambda_2=1.4$. (e, f) $U_2$ vs $\lambda_2$ near the two critical points for $\lambda_1=4$. The insets show the extrapolations of the corresponding critical points.} 
\end{figure}

\subsection{Full ground-state phase diagram\label{C}}

We have already analyzed and confirmed the phases and phase transitions in the cases of $\lambda_1=0$ and $\lambda_2=0$. To complete the remaining phase diagram, we continue to apply the finite-size scaling method to extrapolate the intersection of Binder cumulants for different sizes to accurately pinpoint the phase boundaries. To demonstrate the reliability of our data, we present a detailed analysis of cases where we vary $\lambda_2$ while fixing $\lambda_1$ at $0.5$,$1.0$, and $4$, as well as the case where we vary $\lambda_1$ while fixing $\lambda_2$ at $1.4$, as presented in Fig.~\ref{fig:FIG.6.}. The overall phase diagram for the SHO lattice is presented in Fig.~\ref{fig:FIG.2.}. Within the chosen parameter range, the SHO lattice model exhibits five distinct phases: the hexagon phase, the OSD phase, the LSD phase, the Haldane-like SPT phase, and the AFM phase. The AFM phase is surrounded by other nonmagnetic phases in the phase diagram.

\begin{figure}[t]
\centering
\includegraphics[width=0.4\textwidth]{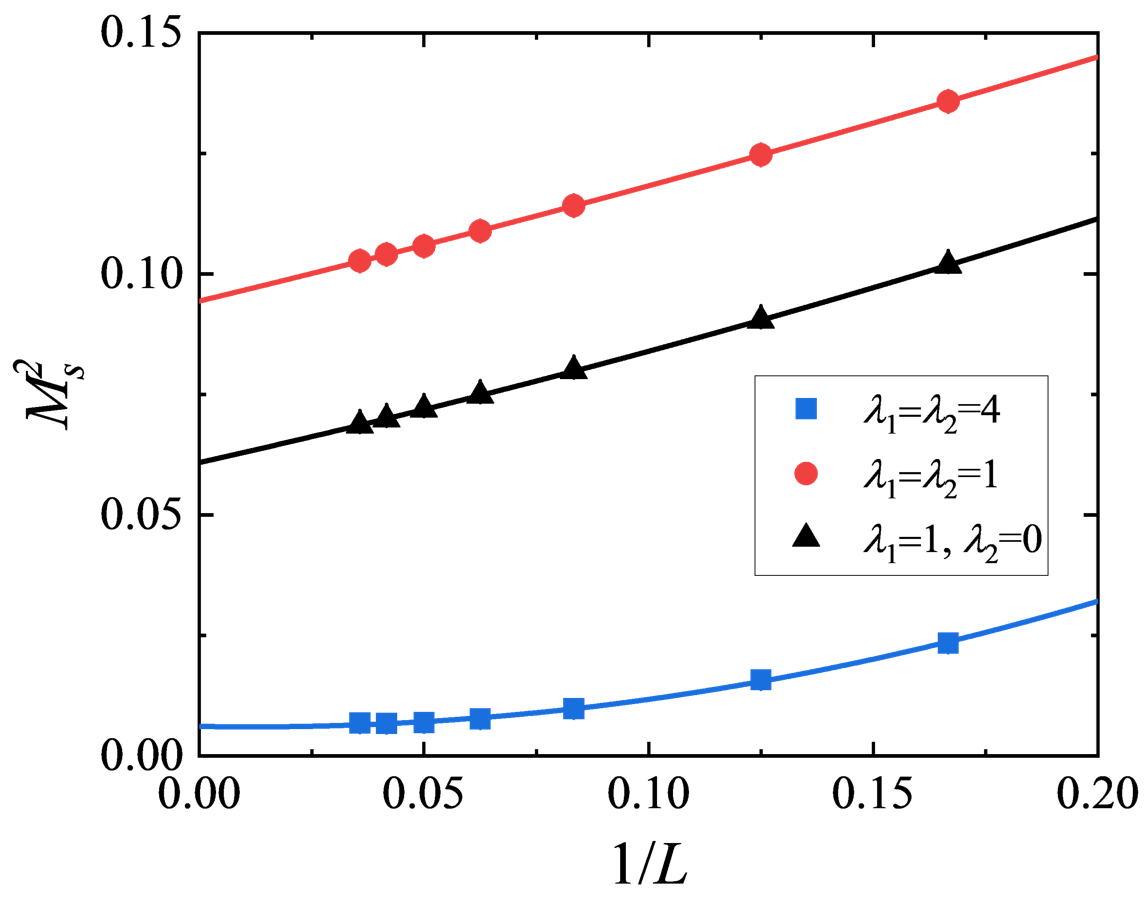}
\caption{\label{fig:FIG.7.}The extrapolation curves of the squared magnetization for different parameter settings $\lambda_1=\lambda_2=4$; $\lambda_1= \lambda_2=1$; and $\lambda_1=1$, $\lambda_2=0$. Their extrapolated values are $M^2_s(\infty) =0.0061(7)$, $0.0944(3)$, and $0.0609(2)$ respectively.} 
\end{figure}

In the vertical path with $\lambda_1$ fixed at $0.5$ and varying $\lambda_2$, two phase transitions occur: one from the hexagon phase to the AFM phase, and the other from the AFM phase to the LSD phase. Through finite-size extrapolation, the critical points are accurately determined to be $\lambda_{2,c} =0.323(7)$ and $2.040(1)$. In the vertical path with $\lambda_1$ fixed at $1$ and varying $\lambda_2$, only one quantum phase transition is observed, namely from the AFM phase to the LSD phase. The critical point for this transition is accurately identified as $\lambda_{2,c}= 2.320(2)$. At the point where $\lambda_1=\lambda_2=1$, the system is topologically equivalent to the square-lattice Heisenberg model. In the thermodynamic limit, the staggered magnetic order is extrapolated to approximately $0.307$ as shown in Fig.~\ref{fig:FIG.7.}, which aligns closely with previously reported results~\cite{sandvikLoopUpdatesVariational2010}. When $\lambda_1=1$ and $\lambda_2=0$, the magnetic order is reduced compared to the case where $\lambda_1=\lambda_2=1$. In the vertical path with $\lambda_1$ fixed at $4$ and varying $\lambda_2$, the AFM phase region shrinks compared to cases with smaller $\lambda_1$ values. Two continuous quantum phase transitions occur between the AFM phase and the OSD phase, as well as between the AFM phase and the LSD phase. The critical points for these transitions are accurately determined to be $\lambda_{2,c}=3.939(6)$ and $4.228(4)$. And the AFM order at the $\lambda_1=\lambda_2=4$ point is very small compared to the other AFM region, as shown in Fig.~\ref{fig:FIG.7.}. To illustrate the quantum phase transition between the Haldane-like SPT phase and the AFM phase, we consider a horizontal path where $\lambda_1$ is varied while $\lambda_2$ is fixed at $1.4$. As shown in Fig.~\ref{fig:FIG.6.}(d), the Binder cumulants for different system sizes exhibit intersections that remain relatively unchanged with increasing system sizes. By performing extrapolation, we accurately determine a quantum critical point at $\lambda_{1,c}=0.106(1)$. This is analogous to the case of coupled Haldane chains transitioning into a 2D square lattice, where a similarly small critical interchain interaction has been reported~\cite{PhysRevB.103.024412}.

We plot the identified phase transition points on the phase diagram and use the same finite-size scaling method to determine additional points across the phase diagram. This process results in a detailed and comprehensive phase diagram for the Heisenberg model on the 2D SHO lattice, as displayed in Fig.~\ref{fig:FIG.2.}. Due to the presence of an even sublattice, the AFM phase tends to emerge readily. It occupies a middle area of the phase diagram and is surrounded by four other nonmagnetic phases. 

\begin{figure}[t]
\centering
\includegraphics[width=\linewidth]{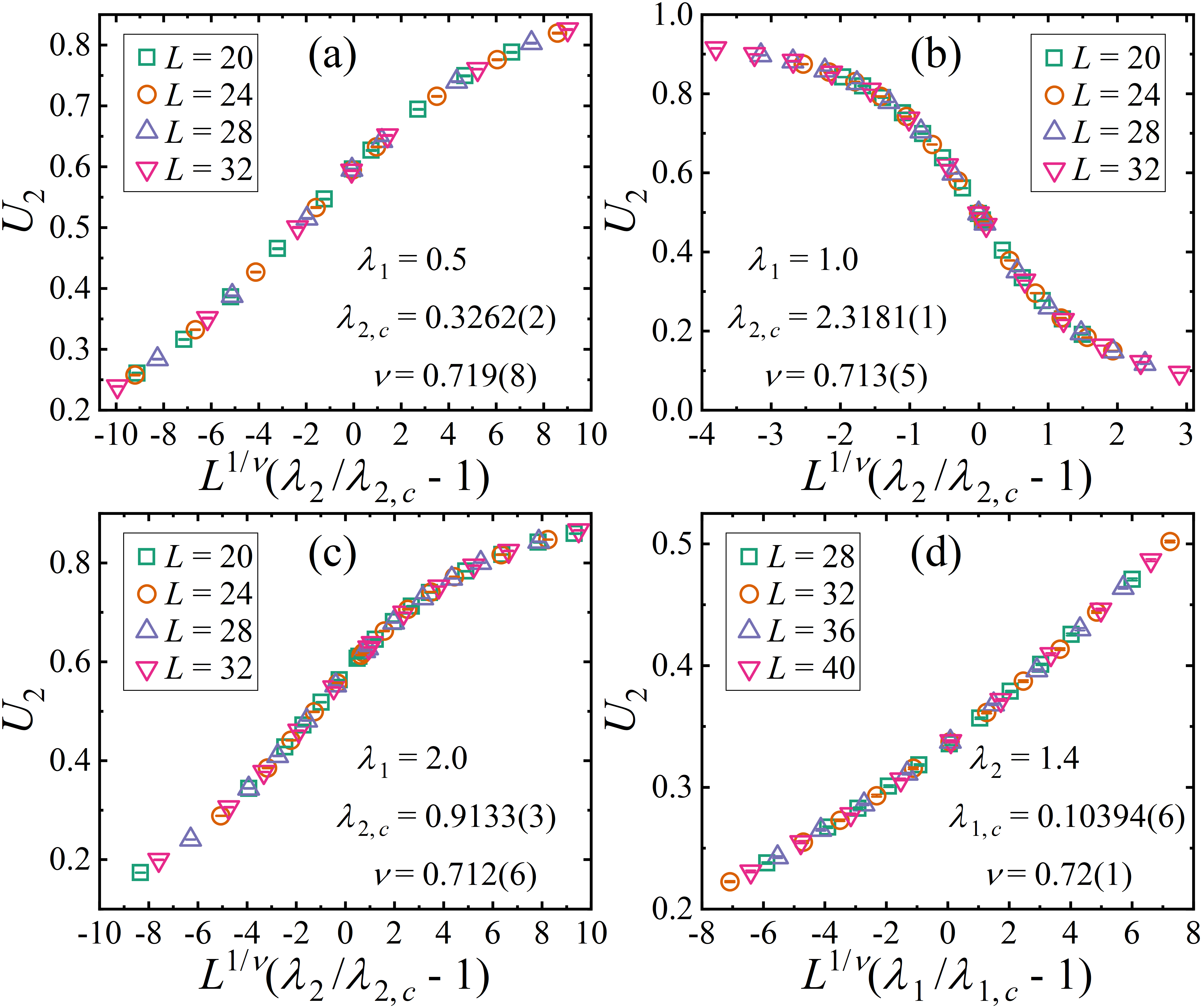}
\caption{(a-d) The data collapse of the Binder cumulant $U_2$ for the phase transitions between the AFM phase and the hexagon phase ($\lambda_1=0.5$), LSD phase ($\lambda_1=1.0$), OSD phase ($\lambda_1=2.0$), and Haldane-like SPT phase ($\lambda_2=1.4$), respectively. Here, $\lambda_{1/2,c}$ and $\nu$ represent the optimal critical point and critical exponent, respectively. The error estimation is based on a 95\% confidence interval of Gaussian distribution.}
\label{fig:FIG.8.}
\end{figure}

\subsection{Critical behaviors\label{D}}
After studying the position of each phase in the model clearly, we continue to explore the critical behavior in the phase transition. Since the antiferromagnetic phase breaks the spin $SU(2)$ continuous symmetry in this model while the other phases such as the dimer phase or hexagon phase do not break this continuous symmetry and translation symmetry, it is reasonable to conjecture that the phase transition of this model should belong to the three-dimensional (3D) $O(3)$ universality class. At present, there are precise critical exponent values for the $O(3)$ universality class in the literature~\cite{PhysRevB.65.144520}. We only need to do the data collapse of the Binder cumulant $U_2$ and compare the obtained critical exponent with the known value for confirmation. The finite-size scaling form of the Binder cumulant is given by $U_2(\lambda_{1/2},L)=f[L^{1/\nu}(\lambda_{1/2}-\lambda_{1/2,c})/\lambda_{1/2,c}]$, where $f$ is the scaling function, $\lambda_{1/2,c}$ is the critical point, and $\nu$ is the critical exponent of correlation length. To extract the optimal critical point and critical exponent from the finite-size Binder cumulant, we employ the scaling analysis method based on Gaussian process regression proposed in Refs.~\cite{harada2011pre,harada2015pre}, which enables data from different system sizes to collapse onto a single smooth scaling function at the optimal best-fit critical parameters. We select representative paths from the four phase boundaries in the phase diagram and apply the above scaling analysis method to determine their critical exponents and judge all phase transition types accordingly.

The results of data collapse are shown in Fig.~\ref{fig:FIG.8.}, where the best estimates of the critical point $\lambda_{1/2,c}$ and critical exponent $\nu$ for different phase transitions are presented. The critical points obtained here are all consistent with the phase transition points shown in Fig.~\ref{fig:FIG.2.}. The critical exponents $\nu$ we obtain here are all in good agreement with the $O(3)$ value $\nu\approx0.7112$~\cite{PhysRevB.65.144520}, indicating that the phase transition from the four nonmagnetic phases to the AFM phase in the Heisenberg model on the SHO lattice indeed belongs to the 3D $O(3)$ universality class.

\subsection{Stability of the Haldane-like SPT phase\label{E}}

Our previous study of the $J_1\text{-}J_3$ ladder revealed a Haldane-like SPT phase, intervening between the hexagon and the LSD phases. This phase was identified through characteristic signatures in both the edge spectrum and the entanglement spectrum.

\begin{figure}[t]
\centering
\includegraphics[width=\linewidth]{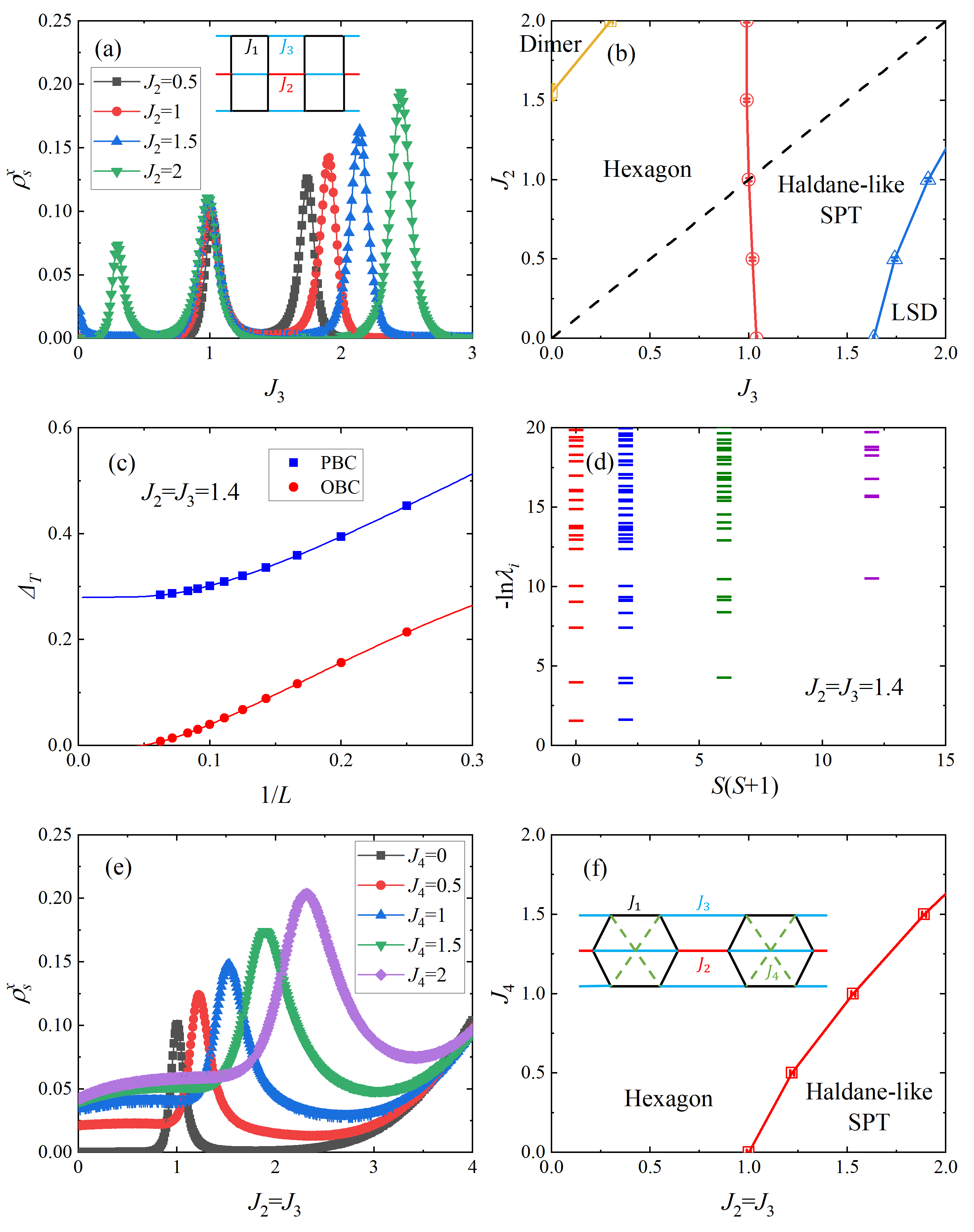}
\caption{(a) Spin stiffness along the $x$ direction as a function of $J_3$ at fixed $J_2$. Peak positions indicate phase boundaries. Inset: Ladder geometry with a rectangular unit cell.
(b) $J_3\text{-}J_2$ phase diagram determined from spin stiffness. A large $J_2$ can induce an inter-unit-cell dimer phase, while introducing $J_2$ expands the region of the Haldane-like SPT phase.
(c) Extrapolation of triplet gaps at $J_2=J_3=1.4$ under PBCs and OBCs.
(d) Entanglement spectrum at $J_2=J_3=1.4$ for system size $L=32$.
(e) Spin stiffness along the $x$ direction as a function of $J_2=J_3$ at fixed $J_4$.
(f) Phase diagram of the $J_3\text{-}J_4$ ladder model with $J_2=J_3$. Inset: Ladder geometry with a hexagonal structure.}
\label{fig:FIG.9.}
\end{figure}

While this model currently exists only in theoretical calculations, its potential realization in magnetic materials highlights the importance of the nearest-neighbor inter-unit-cell antiferromagnetic exchange interaction, $J_2$ [denoted by the red lines in the inset of Fig.~\ref{fig:FIG.9.}(a)], which cannot be neglected. This leads us to investigate the effect of $J_2$ on the SPT phase. Based on the phase diagram obtained from QMC-calculated spin stiffness for a system size of $L=24$, shown in Figs.~\ref{fig:FIG.9.}(a) and \ref{fig:FIG.9.}(b), we find that $J_2$ does not destroy the SPT phase. On the contrary, it enlarges the region of the SPT phase—a promising result for material realization. To further confirm the SPT phase at $J_2=J_3=1.4$, we also present the triplet gaps under PBCs and OBCs, as well as the entanglement spectrum, in Figs.~\ref{fig:FIG.9.}(c) and \ref{fig:FIG.9.}(d). The emergence of end states and a twofold degenerate entanglement spectrum confirms the presence of the SPT phase.

If a material's lattice is built on a hexagonal unit cell and $J_3$ is non-negligible, other third-neighbor interactions within the hexagon—such as $J_4$—should also be taken into account. To examine the effect of $J_4$, we select a diagonal path with $J_2=J_3$ in Fig.~\ref{fig:FIG.9.}(b) and employ spin stiffness to map out the $J_4$ phase diagram, as shown in Figs.~\ref{fig:FIG.9.}(e) and \ref{fig:FIG.9.}(f). Our finite-size numerical results indicate that for the SPT phase to be preserved, these third-neighbor interactions $J_4$ should generally be weaker than $J_3$. Additionally, next-nearest-neighbor interactions within the hexagon constitute another possible perturbation. Although such frustrating interactions may influence the SPT phase, it can remain stable as long as they are sufficiently weak due to the finite topological gap of the SPT phase.

Finally, we note that the preceding discussion on SPT stability is based on the ladder geometry. When extended to two dimensions, our phase diagrams in Fig.~\ref{fig:FIG.2.} indicate that weak inter-ladder couplings ($J_2\approx 0.1J_1$) can easily destabilize the one-dimensional SPT phase, driving a transition into a magnetically ordered phase.

\section{CONCLUSIONS\label{CONCLUSIONS}}
In this paper, we have investigated the ground-state properties and critical behavior of the $S=1/2$ Heisenberg model on the SHO lattice. The SHO lattice, characterized by a polygonal nested structure with periodically arranged four-, six-, and eight-membered rings on a two-dimensional plane, offers a rich landscape for exploring quantum magnetic phenomena. To study the rich ground-state phases of the Heisenberg model on the SHO lattice, we introduced three types of interactions: two nearest-neighbor interactions $J_1$ and $J_2$ and a selected third-nearest-neighbor interaction $J_3$ along the $x$ direction. Our numerical approach combined the strengths of SSE-QMC and DMRG methods, with results analyzed through finite-size scaling and extrapolation techniques. This comprehensive analysis enabled us to construct a detailed phase diagram of the ground state, identifying five distinct phases: the AFM phase, hexagon phase, OSD phase, LSD phase, and Haldane-like SPT phase. We focused on representative paths within the phase diagram to present and analyze key data. Especially for the Haldane-like SPT phase, we also employed the DMRG method to identify its edge states and topological properties. By examining the degeneracy of the ground-state energy levels and analyzing the entanglement spectrum under open boundary conditions, we were able to characterize the unique nature of this phase. To connect our findings to real materials, we also investigated the robustness of the SPT phase against other possible interactions. Moreover, by employing the finite-size scaling analysis method, we obtained the critical exponent $\nu$ for the phase transitions in the SHO lattice. Comparison with the known critical exponent of the 3D $O(3)$ universality class revealed that all phase transitions from nonmagnetic phases to the AFM phase in the SHO lattice Heisenberg model belong to this universality class.

Our work, with the Haldane-like SPT phase as one of the important results, significantly advances the theoretical understanding of topological states in quantum spin systems. This finding is of fundamental importance as it provides a platform for exploring complex phase behavior and delivers robust, precise numerical results on a nontrivial lattice.

\begin{acknowledgments}
We would like to thank Zheng-Xin Liu for the helpful discussions. This project is supported by NSFC-12474248, NKRDPC-2022YFA1402802, NSFC-92165204, NSFC-12494591, Leading Talent Program of Guangdong Special Projects (Grant No. 201626003), Guangdong Basic and Applied Basic Research Foundation (Grant No. 2023B1515120013), Guangdong Provincial Key Laboratory of Magnetoelectric Physics and Devices (No. 2022B1212010008), Guangdong Fundamental Research Center for Magnetoelectric Physics (Grant No. 2024B0303390001), and Guangdong Provincial Quantum Science Strategic Initiative (Grant No. GDZX2401010).
\end{acknowledgments}

\bibliography{references}
\end{document}